# Unveiling an in-plane Hall effect in rutile RuO$_2$ films

Meng Wang[1,2,*], Jianbing Zhang[3], Di Tian[3], Pu Yu[3], Fumitaka Kagawa[1,4,*]

[1]*RIKEN Center for Emergent Matter Science (CEMS), Wako, 351-0198, Japan.*
[2] *School of Integrated Circuits and Electronics, MIIT Key Laboratory for Low-Dimensional Quantum Structure and Devices, Beijing Institute of Technology, Beijing, 100081, China*
[3]*State Key Laboratory of Low Dimensional Quantum Physics and Department of Physics, Tsinghua University, Beijing, 100084, China.*
[4]*Department of Physics, Tokyo Institute of Technology, Tokyo 152-8551, Japan.*
[*]*Corresponding authors. Email: wangmeng@bit.edu.cn; kagawa@phys.sci.isct.ac.jp*

**Abstract**

**The in-plane-magnetic-field-induced Hall effect (IPHE) observed in Weyl semimetals and *PT*-symmetric antiferromagnets has attracted increasing attention, as it breaks the stereotype that the Hall effect is induced by an out-of-plane magnetic field or magnetization. To date, the IPHE has been discussed mainly for materials with low-symmetry crystal/magnetic point groups. Here, we show that even if symmetry forbids an inherent IPHE that arises from any mechanism, an apparent IPHE can be generated by selecting a low-symmetry crystalline plane for measurement. For rutile RuO$_2$, although its high symmetry forbids an inherent IPHE, films grown along the low-symmetry (1 1 1) and (1 0 1) orientations are found to exhibit a distinct IPHE. The in-plane Hall coefficients are quantitatively reproduced by referring to the out-of-plane Hall coefficients measured for the high-symmetry (1 0 0) and (0 0 1) planes, indicating that the observed IPHE is caused by a superposition of inequivalent out-of-plane Hall effects. Similar behaviour is also observed for paramagnetic rutile systems, indicating the ubiquity of the apparent IPHE in electronic and spintronic devices with low-symmetry crystalline planes.**

**Introduction**

The Hall effect has been extensively employed in electronic and spintronic devices to study the carrier density, spin-torque effects, etc.[1-9] A longstanding assumption is that the Hall effect generally appears only under a magnetic field ($B$) or magnetization ($M$) along the out-of-plane direction. However, although the topic may not have received much attention, in-plane magnetic fields were reported to be able to generate $B$-odd and $B$-linear Hall voltages in the 1980s, for example, in monoclinic semimetallic $SrAs_3$.[10,11] Recently, the in-plane Hall effect (IPHE), which refers to the $B$-odd (or $M$-odd) Hall voltage induced linearly or nonlinearly by an in-plane $B$ (or $M$), has attracted renewed interest in terms of the topological nature of the band structure, and it has been observed in the inversion-symmetry-broken Weyl semimetals $ZrTe_5$,[12,13] $Pb_{1-x}Sn_xTe$,[14] the $\mathcal{PT}$-symmetric antiferromagnetic $VS_2$-$VS$ superlattice,[15,16] canted antiferromagnetic DyPtBi,[17] and ferromagnetic iron and nickel.[18] The microscopic mechanisms of the IPHEs in these systems has been attributed to the field-induced Berry curvature or the octupole in the magnetization space.[19-21] Very recently, the unexpected IPHE has also been observed in rutile $RuO_2$ films,[22] and the microscopic mechanism has been attributed to the Lorentz force.

Symmetry considerations provide valuable insights into a system to judge whether the phenomenon of interest, regardless of its microscopic origin, can be observed. This is also the case for the IPHE, and the symmetry constraints for allowing the $B$-linear IPHE and $B$-nonlinear IPHE are given in the literature.[16,19-21] In theories, symmetry considerations based on the point group (or magnetic point group) of a material are often examined in the conventional Cartesian coordinate system, in which at least one of the orthonormal bases corresponds to the principal axis of a crystal. In Hall resistivity measurements, however, the Cartesian coordinate system defined in the experiment depends on the crystal plane used in the measurement, which does not always match that used in point-group-based considerations. Therefore, even under the point group for which the IPHE is concluded to be forbidden, an apparent IPHE, which is represented by a superposition of inequivalent out-of-plane Hall effects, may be observed in actual experiments. To explore an inherent IPHE that cannot be represented by the superposition of the out-of-plane Hall effects of a material, having a good understanding of the apparent IPHE is therefore important.

In this work, we investigate the crystal-plane-controlled apparent IPHE in rutile $RuO_2$ films and reveal its symmetry relationship to the out-of-plane Hall effects measured for high symmetry (1 0 0) and (0 0 1) planes. We show that its relatively high-symmetry point group does not allow an inherent IPHE, whereas an apparent IPHE emerges when a Hall-bar device with a low-symmetry crystal plane is made. The in-plane Hall coefficients can be quantitatively reproduced by referring to the ordinary Hall coefficients obtained for the high-symmetry planes, indicating that the apparent IPHE can be understood as a superposition of two inequivalent out-of-plane Hall effects.

## Results

**Symmetry considerations for the IPHE**

Before the experimental results are presented, an overview of the symmetry arguments is provided. When considering the $B$-odd Hall resistivity, $\rho_{zy}^{\text{anti}}$, $\rho_{xz}^{\text{anti}}$, and $\rho_{yx}^{\text{anti}}$ are of interest, where $\rho_{ij}^{\text{anti}}$ ($i, j = x, y, z$) represents the antisymmetric part of the resistivity tensor. By defining the Hall-resistivity vector, which is axial, as $\boldsymbol{\rho}_{\text{Hall}} = {}^t(\rho_{zy}^{\text{anti}}, \rho_{xz}^{\text{anti}}, \rho_{yx}^{\text{anti}})$, the Hall effect can thus be described by $\boldsymbol{E}_{\text{Hall}} = \boldsymbol{\rho}_{\text{Hall}} \times \boldsymbol{J}$, where $\boldsymbol{E}_{\text{Hall}}$ is the Hall electric field perpendicular to the current-density vector $\boldsymbol{J}$.[16, 19-21] Thus, for a Hall-bar device with an $xy$ measurement plane, the presence or absence of a Hall voltage is equivalent to considering whether the $z$-component of $\boldsymbol{\rho}_{\text{Hall}}$ is finite. In this study, we consider collinear antiferromagnets or paramagnets such that $\boldsymbol{\rho}_{\text{Hall}} = \mathbf{0}$ under zero magnetic field (i.e., $\boldsymbol{B} = \mathbf{0}$) and the Hall resistivity is linearly induced in proportion to $\boldsymbol{B}$. Microscopically, a finite $\boldsymbol{\rho}_{\text{Hall}}$ under a finite $\boldsymbol{B}$ is naturally expected considering the field-induced magnetic moment in collinear antiferromagnets or the Lorentz force effect in paramagnets. As long as the $B$-linear response is considered, $\boldsymbol{\rho}_{\text{Hall}}$ can be described by:[16,19]

$$\rho_{\text{Hall}} = R^H B, \qquad (1)$$

where $R^H_{ij} \equiv \partial \rho_{\text{Hall},i}/\partial B_j$ is a rank-two Hall-resistivity-coefficient tensor, which is polar and respect the symmetry of the point group (or magnetic point group in antiferromagnets) under $B = 0$.[16,19] Equation (1) indicates that a $B$-linear $\rho_{\text{Hall},z}$ (= $\rho_{yx}^{\text{anti}}$) can appear under an in-plane field, $B_x$ or $B_y$, if $R^H$ has nonzero off-diagonal components, $R^H_{zx}$ or $R^H_{zy}$, respectively.[16] Whether nonzero off-diagonal components in $R^H$ are allowed is determined by the symmetry of the system and, of course, the choice of the orthonormal bases used for the tensor representation. In particular, the latter factor plays a key role in the emergence of the apparent IPHE in a system with relatively high symmetry.

Symmetry considerations for $R^H$ provide an unambiguous answer to whether an inherent and/or apparent IPHE is observable for a given experimental configuration; below, we call the Cartesian coordinate system defined in the experiment the $L$-frame. Once the Laue (for paramagnets) or magnetic (for magnets) point group of the material under consideration is known, the $R^H$ representation is determined following Neumann's principle, as shown in the literature.[16] Note that this standard representation of the tensor is given in the Cartesian coordinate system with orthonormal bases $(\hat{e}_1^P, \hat{e}_2^P, \hat{e}_3^P)$ ($\hat{e}_3^P$ is chosen to be parallel to the crystal principal axis); below, we call this frame the $P$-frame. If the off-diagonal components of $R^H$ are finite in the $P$-frame, then the IPHE is allowed by symmetry, which we call the inherent IPHE and is the subject addressed in previous studies.[12-18] However, when considering experiments on a Hall-bar device with an arbitrary crystal plane and in-plane axes, the representation in the $L$-frame with orthonormal bases $(\hat{e}_1^L, \hat{e}_2^L, \hat{e}_3^L) = (\hat{e}_x, \hat{e}_y, \hat{e}_z)$ is more relevant.

Thus, in Hall-resistivity experiments, one should consider the $R^H$ representation in the $L$-frame. We introduce the orthogonal matrix $T$ that connects the two basis sets as

$\hat{e}_i^L = \sum_j \hat{e}_j^P T_{ji}$ ($i, j = 1, 2, 3$), or more explicitly:

$$(\hat{e}_1^L, \hat{e}_2^L, \hat{e}_3^L) = (\hat{e}_1^P, \hat{e}_2^P, \hat{e}_3^P) \begin{pmatrix} T_{11} & T_{12} & T_{13} \\ T_{21} & T_{22} & T_{23} \\ T_{31} & T_{32} & T_{33} \end{pmatrix}. \qquad (2)$$

Thus, the $\boldsymbol{R^H}$ representation in the $L$-frame, $R_{ij}^{H'}$, is obtained by referring to that in the $P$-frame, $R_{ij}^H$, as follows:

$$R_{ij}^{H'} = \sum_{(p,q)=(1,2,3)} T_{ip}^{-1} R_{pq}^H T_{qj}. \qquad (3)$$

For an arbitrary-direction $\boldsymbol{B}$ and $x$-direction current $\boldsymbol{J} = J_x \hat{\mathbf{e}}_\mathbf{x}$, the $y$-component of $\boldsymbol{E}_{\text{Hall}}$ (i.e., the Hall electric field measured in the Hall bar) and the corresponding Hall resistivity $\rho_{yx}$ are then given by:

$$\boldsymbol{E}_{\text{Hall}} \cdot \hat{\boldsymbol{e}}_{\boldsymbol{y}} = j_x \left( R_{zx}^{H'} B_x + R_{zy}^{H'} B_y + R_{zz}^{H'} B_z \right) \qquad (4)$$

and:

$$\rho_{yx}^{anti} = R_{zx}^{H'} B_x + R_{zy}^{H'} B_y + R_{zz}^{H'} B_z, \qquad (5)$$

where the first two terms represent the IPHE (i.e., the Hall resistivity due to the in-plane magnetic field, $B_x$ and $B_y$), and the last term represents the out-of-plane Hall effect (i.e., the Hall resistivity due to the out-of-plane magnetic field, $B_z$). Specifically, Eq. (5) explicitly shows that for the IPHE to be observed, either $R_{zx}^{H'}$ or $R_{zy}^{H'}$ should be finite.

In cubic systems, the off-diagonal components of $\boldsymbol{R^H}$ in the $P$-frame are all zero, and all the diagonal components are equivalent $\left( R_{11}^H = R_{22}^H = R_{33}^H \right)$.[16,21] Due to the isotropic nature of the tensor, the off-diagonal components of $\boldsymbol{R^H}$ are all zero in an arbitrary $L$-frame, forbidding both the inherent and apparent $B$-linear IPHEs. In the literature, for the IPHE to be finite under an in-plane $B$ along the $x$ (or $y$) direction, the rotation and mirror symmetry with respect to the $x$- (or $y$-) and $z$-axes must be simultaneously broken.[19-22] The symmetry constraints discussed in the literature are certainly correct for general IPHEs, including "the $B$-nonlinear IPHE", such as $\rho_{yx}^{anti} \propto B^3$, but for the $B$-linear IPHE, which is more relevant in the low-field regime, more rigorous symmetry considerations can be made by referring to $\boldsymbol{R^H}$. In fact, in a cubic Nb-doped SrTiO$_3$ (2 1 0) film, for example, there is no mirror or rotation symmetry with respect to the $x$ // [1 –2 0] and $z$ // [2 1 0] axes, and thus the $B$-nonlinear IPHE is allowed. However, the $B$-linear IPHE is forbidden by the zero off-diagonal components of $\boldsymbol{R^H}$, as experimentally confirmed under $\boldsymbol{B}$ // $x$ (Supplementary Fig. 1). This observation indicates that when considering the $B$-linear IPHE, the symmetry considerations for $\boldsymbol{R^H}$ have a higher resolution. For crystalline systems with lower symmetry than cubic systems, the inherent IPHE may be allowed. Note that, as shown below, even when off-diagonal components of $\boldsymbol{R^H}$ are absent in the $P$-frame (i.e., the inherent $B$-linear IPHE is forbidden by symmetry), the apparent $B$-linear IPHE may be observed if an appropriate $L$-frame is selected to have finite off-diagonal components, $R_{zx}^{H'}$ or $R_{zy}^{H'}$. In the following, such specific examples are given.

**Application of symmetry considerations to RuO$_2$**

Our target system in this study, which explores the IPHE, is the tetragonal rutile RuO$_2$ (lattice constants $a = b \approx \sqrt{2}\,c$),[6-8, 22-28] which has a Laue point group of 4/$mmm$. If this material is antiferromagnetic, as reported in the literature, then the magnetic point group is 4'/$mm'm$.[28] Although whether RuO$_2$ is an antiferromagnet has recently been questioned,[23,29] fortunately, the $\boldsymbol{R^H}$ representation is identical in both cases, with $R_{11}^H = R_{22}^H \neq R_{33}^H$ and $R_{ij}^H = 0$ for $i \neq j$ in the $P$-frame:[16]

$$\boldsymbol{R^H} = \begin{pmatrix} R_{11}^H & 0 & 0 \\ 0 & R_{11}^H & 0 \\ 0 & 0 & R_{33}^H \end{pmatrix} \quad (P\text{-frame}).$$

Thus, the following discussion is not affected by this controversial issue. The absence of off-diagonal components forbids the inherent $B$-linear IPHE. The apparent $B$-linear IPHE is also forbidden for films grown along the principal axis [0 0 1] and the high-symmetry axis [1 0 0] because both $R_{zx}^{H'}$ and $R_{zy}^{H'}$ are zero in each corresponding $L$-frame (Supplementary Note 1). In contrast, in the (1 1 1) film, where the bases of the $L$-frame are $\hat{e}_x$ // $[-1\ -1\ \sqrt{2}]_L$ (i.e., the $[-1\ -1\ 2]$ crystalline axis), $\hat{e}_y$ // $[1\ -1\ 0]_L$ (i.e., the $[1\ -1\ 0]$ crystalline axis), and approximately $\hat{e}_z$ // $[1\ 1\ \sqrt{2}]_L$ (i.e., the $[1\ 1\ 2]$ crystalline axis) (**Fig. 1a**), the $\boldsymbol{R^H}$ representation in the $L$-frame can have off-diagonal components as follows:

$$\boldsymbol{R^H} = \begin{pmatrix} R_{xx}^{H'} & R_{xy}^{H'} & R_{xz}^{H'} \\ R_{yx}^{H'} & R_{yy}^{H'} & R_{yz}^{H'} \\ R_{zx}^{H'} & R_{zy}^{H'} & R_{zz}^{H'} \end{pmatrix} = \begin{pmatrix} \frac{1}{2}(R_{11}^H + R_{33}^H) & 0 & \frac{1}{2}(-R_{11}^H + R_{33}^H) \\ 0 & R_{11}^H & 0 \\ \frac{1}{2}(-R_{11}^H + R_{33}^H) & 0 & \frac{1}{2}(R_{11}^H + R_{33}^H) \end{pmatrix}$$

($L$-frame). (6)

In the $L$-frame, $R_{zx}^{H'}$ is finite, whereas $R_{zy}^{H'}$ is zero. This representation hence indicates that for $\boldsymbol{B}$ // $y$, the field-induced $\boldsymbol{\rho}_{\text{Hall}} = {}^t(0,\ R_{11}^H B_y,\ 0)$ has no z-component (**Fig. 1b**), whereas for $\boldsymbol{B}$ // $x$, the field-induced $\boldsymbol{\rho}_{\text{Hall}} = {}^t\left(\frac{1}{2}(R_{11}^H + R_{33}^H)B_x, 0, \frac{1}{2}(-R_{11}^H + R_{33}^H)B_x\right)$ has a finite z-component (**Fig. 1c**). The $\rho_{yx}^{anti}(\boldsymbol{B})$ (i.e., the z-component of $\boldsymbol{\rho}_{\text{Hall}}$) for an arbitrary-direction $\boldsymbol{B}$ is thus given by:

$$\rho_{yx}^{anti}(\boldsymbol{B}) = R_{zx}^{H'} B_x + R_{zz}^{H'} B_z$$

$$= \frac{1}{2}(-R_{11}^H + R_{33}^H)B_x + \frac{1}{2}(R_{11}^H + R_{33}^H)B_z. \quad (7)$$

Equation (7) explicitly indicates that (i) the $B$-linear IPHE is observed under an in-plane magnetic field along the $x$-axis (the $[-1\ -1\ 2]$ crystal axis) and, more importantly, (ii) the $B$-linear IPHE is a result of a superposition of inequivalent out-of-plane Hall effects. Thus, the $B$-linear IPHE observed under a point group for which the inherent $B$-linear IPHE is forbidden is essentially the same as the out-of-plane Hall effect.

Similarly, in the (1 0 1) film, where the bases of the $L$-frame are $\hat{e}_x$ // $[1\ 0\ -1/\sqrt{2}]_L$ (i.e., the $[1\ 0\ -1]$ crystalline axis), $\hat{e}_y$ // $[0\ 1\ 0]_L$ (i.e., the $[0\ 1\ 0]$ crystalline axis), and approximately $\hat{e}_z$ // $[1\ 0\ \sqrt{2}]_L$ (i.e., the $[1\ 0\ 2]$ crystalline axis) (Supplementary Fig. 2a), an apparent IPHE can also appear (see Supplementary Note 1 for details). For the (1 0 1) film, $\rho_{yx}^{anti}(\boldsymbol{B})$ for an arbitrary-direction $\boldsymbol{B}$ is given by:

$$\rho_{yx}^{anti}(\boldsymbol{B}) = \frac{\sqrt{2}}{3}(R_{11}^H - R_{33}^H)B_x + \left(\frac{1}{3}R_{11}^H + \frac{2}{3}R_{33}^H\right)B_z, \qquad (8)$$

indicating that the apparent IPHE is observed under the in-plane magnetic field along the *x*-axis (the [1 0 -1] crystalline axis). For brevity, superscript "anti" is omitted below.

**Absence of the IPHE in RuO₂ (1 0 0) and (0 0 1) films**

To experimentally verify the *L*-frame-dependent apparent IPHE, we fabricated rutile $RuO_2$ films on insulating $TiO_2$ substrates with different orientations via pulsed laser deposition (PLD). The thicknesses of all the films were controlled to be 15–25 nm. The single phase and high quality of the films were confirmed by X-ray diffraction (XRD) measurements, as shown in Supplementary Figs. 3a–d. Then, transport measurements were carried out with Hall bars fabricated in each film (see the Methods section for details). The longitudinal resistivity ($\rho_{xx}$) of all the films shows a metallic temperature dependence with small differences (Supplementary Fig. 4a), which can be attributed to the crystalline anisotropy and different strain-induced defects. In our Hall resistivity measurements of $RuO_2$ thin films with various growth orientations, no Hall voltage is observed under zero field within the experimental resolution, justifying the application of Eq. (1). This observation also suggests, if anything, that our $RuO_2$ films do not have antiferromagnetism, as previously reported (**Supplementary Note 2**).[23,29]

**Figure 2a** shows the magnetic-field-dependent Hall resistivity ($\rho_{yx}$) measured in a $RuO_2$ (1 0 0) film at 50 K. $\rho_{yx}$ shows *B*-linear behavior for the field along the out-of-plane direction. When the magnetic field is applied within the film plane either along the [0 1 0] or [1 0 0] axis, $\rho_{yx}$ is almost zero even at 6 T, indicating the absence of a *B*-linear IPHE. The same results are qualitatively observed for the $RuO_2$ (0 0 1) film (**Fig. 2b**). These results are consistent with the fact that the $\boldsymbol{R^H}$ representation in the corresponding *L*-frame has no off-diagonal component (for more details, see Supplementary Note 1).

**Temperature dependence of $R_{11}^H$ and $R_{33}^H$**

As demonstrated by the symmetry considerations for $R_{11}^H$ and $R_{33}^H$, the out-of-plane Hall coefficients in the (1 0 0) and (0 0 1) films, respectively, are the fundamental quantities that describe the Hall response in any crystal plane under an arbitrary magnetic field direction. $R_{11}^H$ and $R_{33}^H$ were obtained from the $B$-linear Hall resistivity under an out-of-plane field configuration, and their temperature ($T$) dependences are shown in **Fig. 2c**. The Hall coefficients of the $RuO_2$ (1 0 0) and (0 0 1) films indicate $p$-type and $n$-type carriers, respectively. Such a field-direction-dependent charge carrier type has rarely been observed in experiments but has been reported for another rutile system $IrO_2$,[30] and $LaRh_6Ge_4$.[31] The different signs between $R_{11}^H$ and $R_{33}^H$ observed in different $IrO_2$ film planes are attributed to the complicated momentum dependence of the group velocity and mass tensor on the folded Fermi surface of the nonsymmorphic rutile structure.[30]

We also note that $R_{11}^H$ and $R_{33}^H$ vary with temperature for both the $RuO_2$ and $IrO_2$ films. The variations between the values at room temperature and at the lowest temperature are similar for the two films, ~0.02 μΩ cm T$^{-1}$.[30] Given that the Hall coefficients in $RuO_2$ and $IrO_2$ are well reproduced by semiclassical transport theory considering the Lorentz force mechanism,[22,30] the observed temperature dependence may be explained by considering the temperature variations of the Fermi distribution function. Whereas a detailed understanding of the $R_{11}^H$–$T$ and $R_{33}^H$–$T$ profiles is beyond the scope of this study, we demonstrate that $\rho_{yx}(T)$ in any crystal plane under an arbitrary magnetic field direction can be reproduced by referring to the $R_{11}^H(T)$ and $R_{33}^H(T)$ profiles. This simplicity originates from the fact that there are only two independent components in the **$R^H$** representation in the $P$-frame.

**Emergence of the IPHE in the RuO₂ (1 1 1) film**

Compared to the $RuO_2$ (1 0 0) and (0 0 1) films, the $RuO_2$ (1 1 1) film (**Fig. 3a**) shows distinct behaviour. As shown in **Fig. 3b**, $\rho_{yx}$ is linear in $B$ for the field along the out-of-plane direction, and it is close to zero when the field is applied along the in-plane [1 −1 0] axis (the $y$-axis). In contrast, a considerable $B$-linear $\rho_{yx}$ emerges when the field is applied along the in-plane [−1 −1 2] axis (the $x$-axis). The $B$-linear behaviour clearly differs from the so-called planar Hall effect, which is B-quadratic.[32-35] $\rho_{yx}$ under the in-plane field is larger than that under the out-of-plane field.

The magnetic field angular dependence of this film was further probed. **Figures 3c-e** show the $\rho_{yx}$ values (5 K, 14 T) measured with the magnetic field rotating within the *zx*, *zy*, and *xy* planes, and the corresponding rotation angles are denoted by α, β and γ, respectively. All of the angular dependence curves are consistently described by Eq. (7), i.e., the Hall coefficient is zero for $B_y$ (// [1 −1 0]), and the observed $\rho_{yx}$ is expressed by a linear combination of the Hall responses for $B_x$ (// [−1 −1 2]) and $B_z$ (// [1 1 1]). Moreover, an identical $\rho_{yx}$ is observed when the current direction is changed from the [−1 −1 2] to [1 −1 0] directions while the Hall voltage is measured along the [1 1 −2] direction (Figs. 3c-e), confirming that the IPHE is determined by the configuration of the crystal plane and the magnetic field direction and is independent of the current direction in a given crystal plane. This observation reflects the fact that $\rho_{yx}^{\text{anti}}$ is invariant with respect to *z*-axis rotation ($\theta_z$), as explicitly shown below:

$$\begin{pmatrix} \cos\theta_z & -\sin\theta_z & 0 \\ \sin\theta_z & \cos\theta_z & 0 \\ 0 & 0 & 1 \end{pmatrix}^{-1} \begin{pmatrix} 0 & -\rho_{yx} & -\rho_{zx} \\ \rho_{yx} & 0 & -\rho_{zy} \\ \rho_{zx} & \rho_{zy} & 0 \end{pmatrix} \begin{pmatrix} \cos\theta_z & -\sin\theta_z & 0 \\ \sin\theta_z & \cos\theta_z & 0 \\ 0 & 0 & 1 \end{pmatrix}$$
$$= \begin{pmatrix} 0 & -\rho_{yx} & -\rho_{zx}\cos\theta_z - \rho_{zy}\sin\theta_z \\ \rho_{yx} & 0 & \rho_{zx}\sin\theta_z - \rho_{zy}\cos\theta_z \\ \rho_{zx}\cos\theta_Z + \rho_{zy}\sin\theta_z & -\rho_{zx}\sin\theta_z + \rho_{zy}\cos\theta_z & 0 \end{pmatrix}.$$
(9)

As long as the antisymmetric part of the resistivity tensor is discussed, $\rho_{yx}$ (= −$\rho_{xy}$) is thus expected to always be independent of the in-plane direction of the current.

### Reproducing the Hall coefficients via $R_{11}^H$ and $R_{33}^H$

To quantitatively verify Eq. (7), checking whether the $\rho_{yx}$ data measured for the (1 1 1) film can be reproduced by referring to $R_{11}^H$ and $R_{33}^H$, including their temperature dependence, is more rigorous. In the RuO$_2$ (1 1 1) film, the out-of-plane Hall coefficient changes with temperature, and the sign is even reversed (**Fig. 4a**), whereas the in-plane Hall coefficient is rather insensitive to temperature (**Fig. 4b**). Thus, determining whether these contrasting and seemingly nontrivial temperature dependences can be reproduced via Eq. (7) and the experimentally determined $R_{11}^H-T$ and $R_{33}^H-T$ profiles is interesting (Fig. 2c). Note that Eq. (7) predicts that the temperature dependences of the out-of-plane and in-plane Hall coefficients in the (1 1 1) film are given by $[R_{11}^H(T) + R_{33}^H(T)]/2$ and $[-R_{11}^H(T) + R_{33}^H(T)]/2$, respectively.

**Figure 4c** shows a comparison of both the out-of-plane and in-plane Hall coefficients in the (1 1 1) film between the experimentally determined profiles and the temperature profiles calculated via Eq. (7). The experimental results are quantitatively reproduced by the calculations based on Eq. (7). The small discrepancies observed in the comparison presumably originate from the slight differences in the lattice constants among the different samples, which is caused by different strain effects depending on the film growth direction (Supplementary Figs. 3e-h). In addition, the results reconstructed using the Hall conductivity [$\sigma_{xy} \approx \rho_{yx} / (\rho_{xx} \rho_{yy} + \rho_{yx}^2)$] show a larger discrepancy with the experimental results (Supplementary Fig. 5) than the results observed in Fig. 4c. This observation suggests that when comparing the out-of-plane and in-plane Hall effects in different $RuO_2$ films, the Hall resistivity, rather than the Hall conductivity, is the appropriate quantity. The $RuO_2$ (1 0 1) film also has a temperature-dependent out-of-plane Hall coefficient and temperature-insensitive in-plane Hall coefficients (Supplementary Fig. 2). The temperature dependences are qualitatively similar to those observed for the $RuO_2$ (1 1 1) film, but the sign of the IPHE is the opposite. The temperature profiles are also reproduced via Eq. (8) (Supplementary Fig. 2).

Given that $\rho_{xx}$ (and thus the scattering time $\tau$) is different for different $RuO_2$ films (Supplementary Fig. 4a), the successful reproduction of the out-of-plane and in-plane Hall coefficients for the $RuO_2$ (1 1 1) and (1 0 1) films via Hall-resistivity coefficients ($R_{11}^H$ and $R_{33}^H$) rather than the Hall conductivity coefficients ($\sigma_{xy}/B$) indicates that the $R_{11}^H$ and $R_{33}^H$ are approximately $\tau$-independent. Such a behavior suggests that the IPHE is likely governed by a $\tau$-insensitive mechanism, such as the Lorentz force.[22]

The recent study on RuO$_2$ thin films also argues that the contribution of the Lorentz force is dominant, but their approach is different:[22] They noted first-principles calculations and the experimental observation that the *B*-linear in-plane Hall resistivity is nearly independent of temperature and the in-plane current direction.

Our analysis demonstrates that a temperature-dependent Hall coefficient measured in an arbitrary crystal plane under an arbitrary magnetic field configuration can be reproduced based on the information of only $R_{11}^H(T)$ and $R_{33}^H(T)$. In our framework, the approximately temperature-independent behaviour of the IPHE is interpreted as reflecting the smallness of $\frac{d}{dT}(-R_{11}^H + R_{33}^H)$. Given the fact that $\frac{d}{dT}R_{11}^H$ and $\frac{d}{dT}R_{33}^H$ themselves are not negligible (Fig. 2c) and that $R_{11}^H$ and $R_{33}^H$ are inequivalent quantities, the weak temperature dependence of the *B*-linear IPHE in RuO$_2$ appears to be coincidental, at least from the symmetry point of view.

**Observation of the IPHE in V$_{0.95}$W$_{0.08}$O$_2$ and MoO$_2$ (111) films**

Below, we show that the apparent IPHE is allowed in paramagnetic materials. Here, we fabricated rutile V$_{0.95}$W$_{0.08}$O$_2$ and MoO$_2$ films on TiO$_2$ (1 1 1) to investigate the IPHE. V$_{0.95}$W$_{0.08}$O$_2$ has a metallic state at temperatures above 200 K, accompanied by a metal-insulator transition analogous to that of VO$_2$,[36-39] whereas MoO$_2$ is metallic down to 2 K [40] (Supplementary Figs. 4b and 4c). The transport results for V$_{0.95}$W$_{0.08}$O$_2$ and MoO$_2$ (1 1 1) are similar, in which the IPHE is observed for an in-plane magnetic field along the [−1 −1 2] direction but not for a field along the [1 −1 0] direction (Supplementary Fig. 6). These observations are also consistent with Eq. (7) and indicate that a specific magnetic order is not a key ingredient of the apparent IPHE.

**Discussion**

In general, the variations in the resistivity tensor due to a magnetic field, $\Delta\rho_{ij}(\boldsymbol{B}) \equiv \rho_{ij}(\boldsymbol{B}) - \rho_{ij}(\boldsymbol{B} = \boldsymbol{0})$, can be expressed as the sum of a symmetric tensor $\Delta\rho_{ij}^{sym}$ and an antisymmetric tensor $\Delta\rho_{ij}^{anti}$. The symmetric and antisymmetric tensors are associated with dissipative magnetoresistance and nondissipative Hall transport, respectively. We have considered $\Delta\rho_{ij}$ within the *B*-linear response and treated systems with time-reversal symmetry under zero magnetic field. In this case, the magnetoresistance is parabolic with respect to the magnetic field, and thus, the *B*-linear $\Delta\rho_{ij}^{sym}$ is zero. Therefore, the considerations of the Hall-resistivity vector and the rank-two Hall-resistivity-coefficient tensor are sufficient to describe the *B*-linear magnetotransport. However, for a system with broken time-reversal symmetry, a *B*-linear magnetoresistance can appear; thus, *B*-linear $\Delta\rho_{ij}^{sym} \neq 0$ not only for the diagonal components but also for the off-diagonal components in an arbitrary Cartesian coordinate system. As a result, the measured $\Delta\rho_{ij}(B)$ is described by the sum of the *B*-linear $\Delta\rho_{ij}^{sym}$ and the *B*-linear $\Delta\rho_{ij}^{anti}$. This should be taken into account when applying the methods used in this study to systems with broken time-reversal symmetry.

## Conclusions

In this work, by controlling the growth orientation along different crystalline axes, we observed IPHEs in rutile $RuO_2$ (1 1 1) and (1 0 1) films. The Hall coefficients in the (1 1 1) and (1 0 1) films, including for both the out-of-plane and in-plane Hall effects, can be quantitatively reproduced by referring to the out-of-plane Hall coefficients of the (1 0 0) and (0 0 1) films, indicating that the observed IPHE originates from a superposition of two inequivalent out-of-plane Hall effects. Therefore, we call this IPHE an apparent IPHE. The existence of the apparent IPHE in paramagnetic rutile $V_{0.95}W_{0.08}O_2$ and $MoO_2$ suggests that it is a generic feature associated with a wide range of materials. The apparent IPHE may be more pronounced than the out-of-plane Hall effect, especially if the out-of-plane Hall coefficients change sign depending on the crystal plane, as in the case of rutile $RuO_2$. We envision that the apparent IPHE may greatly expand the design flexibility of electronic and/or spintronic devices that utilize the Hall voltage and Hall current.[7,8,41-43]

## Methods

### Film fabrication

All of the films were fabricated on insulating TiO$_2$ substrates via a PLD system.[44] RuO$_2$ films of various orientations were grown at 300 °C, with an oxygen pressure of 2.5 Pa and the laser (248 nm) energy fluence controlled at 1.5 J/cm$^2$. A custom-made target was used, which was fabricated by sintering mixed Ru and RuO$_2$ powders (molar ratio of 1:1) at 1000 °C for 20 hours. The V$_{0.95}$W$_{0.08}$O$_2$ film was grown at 300 °C and 0.5 Pa, with a fluence of 1.5 J/cm$^2$, utilizing a stoichiometric target. The MoO$_2$ film was grown at 400 °C and 10$^{-3}$ Pa, with a fluence of 1.0 J/cm$^2$, utilizing a MoO$_2$ target.

### XRD measurements

XRD was carried out with a high-resolution system (Smartlab, Rigaku) with Cu-K$_{\alpha 1}$ radiation ($\lambda = 0.15406$ nm) at RIKEN and Tsinghua University.

### Hall-bar fabrication and transport measurements

Hall bars with a size of 300 μm × 60 μm were fabricated on the RuO$_2$ films through photolithography and an Ar/O-ion plasma etching system. The beam fluxes of Ar and O ions were set as 15 and 1.5 sccm, respectively, to avoid the introduction of oxygen vacancies into the substrate. The longitudinal resistance and Hall resistance were measured with a DC current of 500 μA for MoO$_2$ and RuO$_2$ and 50 μA for V$_{0.92}$W$_{0.08}$O$_2$ using a 14 T PPMS with a rotator (Quantum Design). The angle-dependent $\rho_{yx}$ was obtained by ($\rho_{yx}$ ($H$) – $\rho_{yx}$ (–$H$)) / 2 to subtract the contribution of the longitudinal resistivity, which is an even function of the magnetic field.

## Data availability

The data that support the plots within this paper are available from the corresponding author upon reasonable request.

## References

1. Ohtomo, A. & Hwang, H. A high-mobility electron gas at the LaAlO$_3$/SrTiO$_3$ heterointerface. *Nature* **427**, 423-426 (2004).

2. Lee, H. et al. Direct observation of a two-dimensional hole gas at oxide interfaces. *Nat. Mater.* 17, 231-236 (2018).

3. Nagaosa, N., Sinova, J., Onoda, S., MacDonald, A. H. & Ong, N. P. Anomalous Hall effect. *Rev. Mod. Phys*. **82**, 1539-1592 (2010).

4. Manchon, A. et al. Current-induced spin-orbit torques in ferromagnetic and antiferromagnetic systems. *Rev. Mod. Phys.* **91**, 035004 (2019).

5. Liu, L. et al. Current-induced magnetization switching in all-oxide heterostructures. *Nat. Nanotech*. **14**, 939-944 (2019).

6. Karube, S. et al. Observation of spin-splitter torque in collinear antiferromagnetic RuO$_2$. *Phys. Rev. Lett.* **129**, 137201 (2022).

7. Bai, H. et al. Efficient spin-to-charge conversion via altermagnetic spin splitting effect in antiferromagnet RuO$_2$. *Phys. Rev. Lett*. **130**, 216701 (2023).

8. Bose, A. et al. Tilted spin current generated by the collinear antiferromagnet ruthenium dioxide. *Nat. Electron.* **5**, 267-274 (2022).

9. Patton, M. et al. Symmetry control of unconventional spin-orbit torques in IrO$_2$. *Adv. Mater*. **35**, 2301608 (2023).

10. Möllendorf, M. and Bauhofer, W. First-order longitudinal Hall effect, *Phys. Rev. B* **30**, 1099(R) (1984).

11. Bauhofer, W. Longitudinal Hall effect in SrAs$_3$, *Phys. Rev. B* **32**, 1183 (1985).

12. Liang, T. et al. Anomalous Hall effect in ZrTe$_5$. *Nat. Phys*. **14**, 451-455 (2018).

13. Wang, Y. et al. Gigantic magnetochiral anisotropy in the topological semimetal ZrTe$_5$. *Phys. Rev. Lett.* **128**, 176602 (2022).

14. Zhang C.-L. et al. Berry curvature generation detected by Nernst responses in ferroelectric Weyl semimetal. *Proc. Natl Acad. Sci. USA* **118**, e2111855118 (2021).

15. Zhou, J. et al. Heterodimensional superlattice with in-plane anomalous Hall effect. *Nature* **609**, 46-51 (2022).

16. Cao, J. et al In-plane anomalous Hall effect in PT-symmetric antiferromagnetic materials. *Phys. Rev. Lett*. **130**, 166702 (2023).

17. Chen, J. et al. Unconventional Anomalous Hall Effect in the Canted Antiferromagnetic Half-Heusler Compound DyPtBi. *Adv. Funct. Mater.* **32**, 2107526 (2022).

18. Peng, W. et al. Observation of the in-plane anomalous Hall effect induced by octupole in magnetization space. https://doi.org/10.48550/arXiv.2402.15741

19. Li, L. et al. Planar Hall effect in topological Weyl and nodal-line semimetals. *Phys. Rev. B* **108**, 085120 (2023).

20. Kurumaji, T. Symmetry-based requirement for the measurement of electrical and thermal Hall conductivity under an in-plane magnetic field. *Phys. Rev. Research* **5**, 023138 (2023).

21. Wang, H. et al. Orbital origin of the intrinsic planar Hall effect. *Phys. Rev. Lett.* **132**, 056301 (2024).

22. Cui, Y. et al. Antisymmetric planar Hall effect in rutile oxide films induced by the Lorentz force. *Sci. Bull.* **69**, 2362 (2024).

23. Hiraishi, M. et al. Nonmagnetic Ground State in $RuO_2$ Revealed by Muon Spin Rotation. *Phys. Rev. Lett.* **132**, 166702 (2024).

24. Šmejkal, L. et al. Anomalous Hall antiferromagnets. *Nat. Rev. Mater.* **7**, 482-496 (2022).

25. Berlijn, T. et al. Itinerant antiferromagnetism in $RuO_2$. *Phys. Rev. Lett.* **118**, 077201 (2017).

26. Šmejkal, L., Sinova, J. & Jungwirth, T. Emerging research landscape of altermagnetism. *Phys. Rev. X* **12**, 040501 (2022).

27. Feng, Z. et al. An anomalous Hall effect in altermagnetic ruthenium dioxide. *Nat. Electron.* **5**, 735-743 (2022).

28. Šmejkal, L., González-Hernández, R., Jungwirth, T. & Sinova, J. Crystal time-reversal symmetry breaking and spontaneous Hall effect in collinear antiferromagnets. *Sci. Adv.* **6**, eaaz8809 (2020).

29. Lovesey S. W. et al. Magnetic structure of $RuO_2$ in view of altermagnetism. *Phys. Rev. B* **108**, L121103 (2023).

30. Uchida, M. et al. Field-direction control of the type of charge carriers in nonsymmorphic $IrO_2$. *Phys. Rev. B* **91**, 241119(R) (2015).

31. Luo, S. et al. Direction-dependent switching of carrier type enabled by Fermi surface geometry. *Phys. Rev. B* **108**, 195146 (2023).

32. Seemann, K. M. et al. Origin of the planar Hall effect in nanocrystalline $Co_{60}Fe_{20}B_{20}$. *Phys. Rev. Lett.* **107**, 086603 (2011).

33. Taskin, A. A. et al. Planar Hall effect from the surface of topological insulators. *Nat. Commun.* **8**, 1340 (2017).

34. Wadehra, N. et al. Planar Hall effect and anisotropic magnetoresistance in polar-polar interface of $LaVO_3$-$KTaO_3$ with strong spin-orbit coupling. *Nat. Commun.* **11**, 874 (2020).

35. Li, Z. et al. Planar Hall effect in $PtSe_2$. *J. Appl. Phys.* **127**, 054306 (2020).

36. Takami, H. et al. Filling-controlled Mott transition in W-doped $VO_2$. *Phys. Rev. B* **85**, 205111 (2012).

37. Okuyama, D. et al. X-ray study of metal-insulator transitions induced by W doping and photoirradiation in $VO_2$ films. *Phys. Rev. B* **91**, 064101 (2015).

38. Qazilbash, M. M. et al. Mott transition in $VO_2$ revealed by infrared spectroscopy and nano-imaging. *Science* **318**, 1750-1753 (2007).

39. Lee, D. et al. Isostructural metal-insulator transition in $VO_2$. *Science* **362**, 1037-1040 (2018).

40. Ma, C.-H. et al. Van der Waals epitaxy of functional $MoO_2$ film on mica for flexible electronics. *Appl. Phys. Lett.* **108**, 253104 (2016).

41. Avci, C. et al. Unidirectional spin Hall magnetoresistance in ferromagnet/normal metal bilayers. *Nat. Phys.* **11**, 570-575 (2015)

42. Lv, Y. et al. Unidirectional spin-Hall and Rashba-Edelstein magnetoresistance in topological insulator-ferromagnet layer heterostructures. *Nat Commun.* **9**, 111 (2018).

43. Jiang, D. et al. Substrate-induced spin-torque-like signal in spin-torque ferromagnetic resonance measurement. *Phys. Rev. Applied* **21**, 024021 (2024).

44. Liu, J. et al. Emergent weak antilocalization and wide-temperature-range electronic phase diagram in epitaxial $RuO_2$ thin film. *J. Phys.: Condens. Matter* **35**, 405603 (2023).

**Acknowledgments**

M. W. and F. K. thank K. Tanaka, R. Arita, M. Uchida, and T. Kurumaji for fruitful discussions. This study was financially supported by JSPS KAKENHI (Grant No. 21H04442). P. Y. acknowledges the support from the National Key R&D Program of China (grant No. 2021YFE0107900) and the National Natural Science Foundation of China (grant No. 52025024). J. Z. acknowledges the support from Postdoctoral Fellowship Program of CPSF (GZB20240380).

**Author contributions**

M. W. and F. K. conceived the project. M. W. fabricated the samples and performed the transport measurements. M. W., J. Z. and D. T. performed the XRD measurements with the help of P. Y. M. W. and F. K. wrote the manuscript. All the authors discussed the results and commented on the manuscript.

**Competing interests**

The authors declare that they have no competing interests.

**Figures**

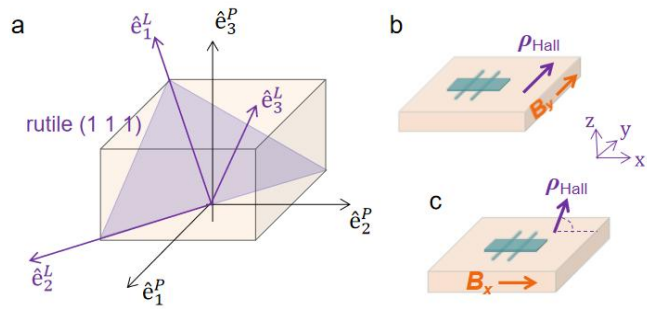

**Figure 1. Schematic illustration of the in-plane Hall effect in rutile (1 1 1) films**. **a**, Illustration of the two different Cartesian coordinate systems, *P*- and *L*-frames. The bases of the *P*-frame, $\{\hat{e}_1^P, \hat{e}_2^P, \hat{e}_3^P\}$, are formed by the orthogonal crystallographic *a*, *b*, and *c* axes, and those of the *L*-frame, $\{\hat{e}_1^L, \hat{e}_2^L, \hat{e}_3^L\} = \{\hat{e}_x, \hat{e}_y, \hat{e}_z\}$, are defined by the (1 1 1) facet, for which transport measurements were performed. $\hat{e}_3^L = \hat{e}_z$ in the *L*-frame is close to the [1 1 2] direction of the rutile structure due to a = b ≈ √2 c. **b, c**, Illustrations of the Hall-resistivity vector $\rho_{\text{Hall}}$ (b) without and (c) with an out-of-plane (oop) component when the magnetic field is applied along the [1 −1 0] (y-axis) and [−1 −1 2] (x-axis) directions, respectively.

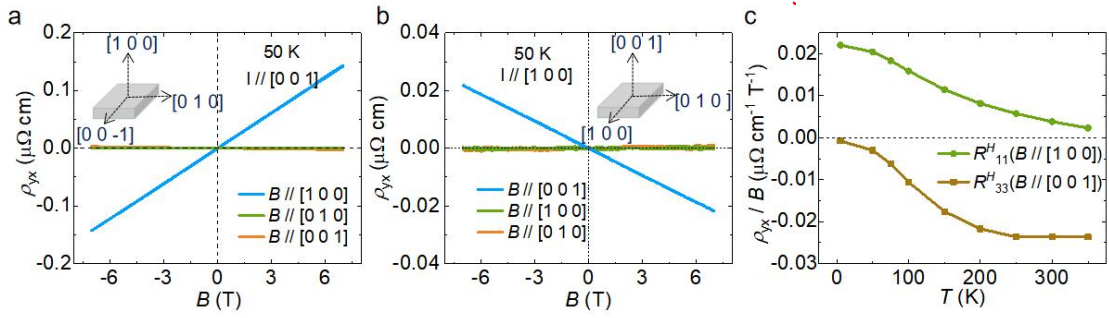

**Figure 2. Absence of the IPHE in high-symmetry RuO$_2$ (1 0 0) and (0 0 1) films**. **a**, **b**, $\rho_{yx}$–$B$ curves measured for the (a) RuO$_2$ (1 0 0), and (b) RuO$_2$ (0 0 1) films with the magnetic field applied in three orthogonal directions. "I" denotes the current. Insets, illustrations of the crystal axes. **c**, Temperature dependence of the Hall coefficients ($\rho_{yx}/B$), i.e., $R^H_{11}$ and $R^H_{33}$, measured with out-of-plane magnetic fields along [1 0 0] and [0 0 1], respectively, for the two films.

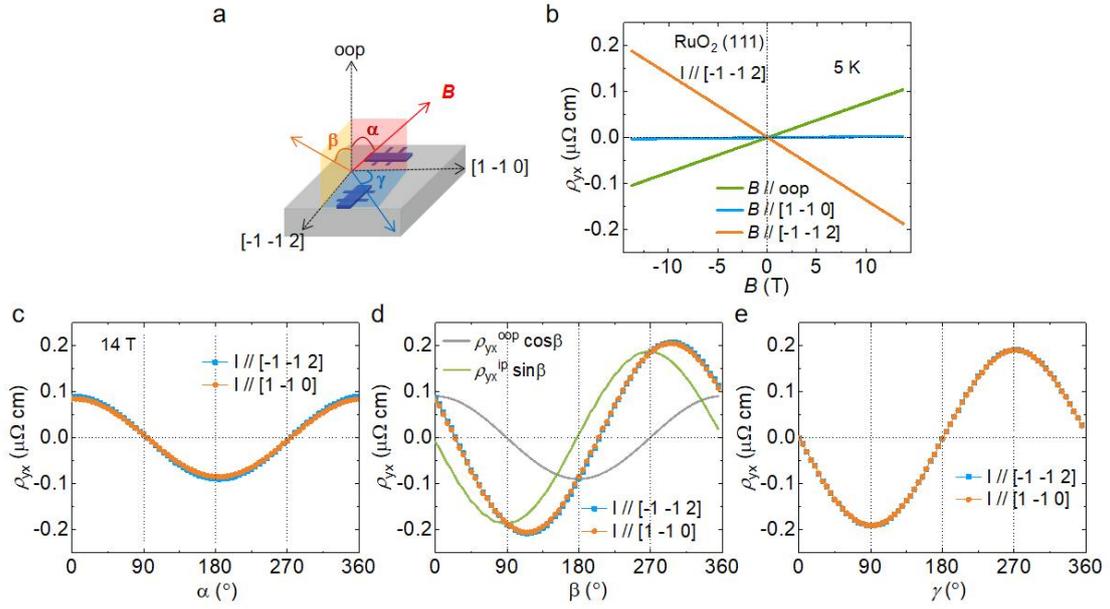

**Figure 3. In-plane Hall effect in a RuO$_2$ (1 1 1) film**. **a**, Schematic illustration of the measurement configuration. $\alpha$, $\beta$ and $\gamma$ denote the rotation angles of the magnetic field (**B**). **b**, $\rho_{yx}$–$B$ curves measured for a RuO$_2$ (1 1 1) film with a magnetic field applied in three directions. The results of the current (I) along [−1 −1 2] are shown. **c**–**e**, Angular dependent $\rho_{yx}$ (5 K, 14 T) with the magnetic field rotating within different planes, as shown in (a). The results for the two orthogonal current directions are shown together for comparison. The data in (c) and (e) show a cos $\alpha$ dependence and a sin $\gamma$ dependence, respectively. The curve in (d) can be well fitted by $\rho_{yx} = \rho_{yx}^{oop} \cos \beta + \rho_{yx}^{ip} \sin \beta$, where $\rho_{yx}^{oop}$ and $\rho_{yx}^{ip}$ denote the Hall resistivities of the magnetic field along the out-of-plane (oop) and in-plane (ip) [−1 −1 2] directions, respectively. The results are well described by Eq. (7), suggesting that an IPHE emerges for the field along the [−1 −1 2] direction, whereas it is forbidden for the field along the [1 −1 0] direction.

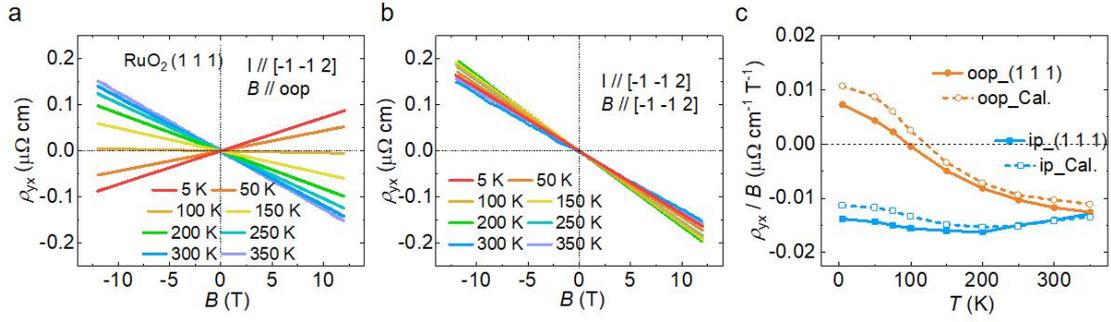

**Figure 4. Temperature-dependent Hall coefficients of RuO$_2$ (1 1 1)**. **a**, **b**, $\rho_{yx}$–$B$ curves at various temperatures measured with a magnetic field along the (a) out-of-plane and (b) in-plane [–1 –1 2] directions. **c**, Hall coefficients (i.e., $\rho_{yx}/B$) measured (solid line) and calculated (dashed line) for RuO$_2$ (1 1 1). The results measured for out-of-plane (oop) and in-plane (ip) magnetic fields are shown together. The calculated data (marked by _Cal.) were obtained via Eq. (7) with the Hall resistivity coefficient data shown in Fig. 2c.